\documentclass[conference]{IEEEtran}
\usepackage{cite}
\usepackage{amsmath,amssymb,amsfonts}
\usepackage{algorithm}
\usepackage{algpseudocode}
\usepackage{graphicx}
\usepackage{subfigure}
\usepackage{textcomp}
\usepackage{xcolor}
\usepackage{multirow} 
\usepackage{mwe}
\usepackage{soul}
\usepackage{subfig}

\setstcolor{red}

\def\BibTeX{{\rm B\kern-.05em{\sc i\kern-.025em b}\kern-.08em
    T\kern-.1667em\lower.7ex\hbox{E}\kern-.125emX}}
    
\usepackage{caption}
\IEEEoverridecommandlockouts

\allowdisplaybreaks

\begin{document}
\bstctlcite{IEEEexample:BSTcontrol}

\title{Tensor Decomposition of Large-scale Clinical EEGs Reveals Interpretable Patterns of Brain Physiology
}
\author{$\text{Teja Gupta}^{1,*}$\thanks{$\text{}^1$ University of Illinois. \{tejag2, nwagh2, scrawal2, varatha2\}@illinois.edu}, $\text{Neeraj Wagh}^{1,*}$, $\text{Samarth Rawal}^{1}$, $\text{Brent Berry}^{2}$\thanks{$\text{}^2$ Mayo Clinic. \{berry.brent, worrell.gregory\}@mayo.edu}, $\text{Gregory Worrell}^{2}$, $\text{Yogatheesan Varatharajah}^{1}$\thanks{$\text{}^*$ These authors contributed equally.}}

\maketitle

\begin{abstract}

Identifying abnormal patterns in electroencephalography (EEG) remains the cornerstone of diagnosing several neurological diseases. The current clinical EEG review process relies heavily on expert visual review, which is unscalable and error-prone. In an effort to augment the expert review process, there is a significant interest in mining population-level EEG patterns using unsupervised approaches. Current approaches rely either on two-dimensional decompositions (e.g., principal and independent component analyses) or deep representation learning (e.g., auto-encoders, self-supervision). However, most approaches do not leverage the natural multi-dimensional structure of EEGs and lack interpretability. In this study, we propose a tensor decomposition approach using the canonical polyadic decomposition to discover a parsimonious set of population-level EEG patterns, retaining the natural multi-dimensional structure of EEG recordings (time$\times$space$\times$frequency). We then validate their clinical value using a cohort of patients with  varying stages of cognitive impairment. Our results show that the discovered patterns reflect physiologically meaningful features and accurately classify the stages of cognitive impairment (healthy vs mild cognitive impairment vs Alzheimer's dementia) with substantially fewer features compared to classical and deep learning-based baselines. We conclude that the decomposition of population-level EEG tensors recovers expert-interpretable EEG patterns that can aid in studying smaller specialized clinical cohorts.
\end{abstract}


\section{Introduction}

EEG measures the electrical activity of the brain \cite{Binnie1308}. The identification of abnormal patterns via expert visual review is the current gold-standard test for diagnosing a wide range of neurological, psychiatric, and sleep disorders using EEGs \cite{ebersole2003current}. However, expert visual review is tedious, unscalable, error-prone, and subject to reviewer bias \cite{agarwal1998automatic}. The recent availability of large population-level clinical EEG datasets, such as the Temple University Hospital EEG corpus (TUH-EEG) \cite{obeid2016temple}, has sparked interest in the data-driven discovery of EEG patterns, to augment the expert review process. 

Nonetheless, the automated discovery of interpretable EEG patterns remains challenging. Despite the availability of large labeled EEG datasets, the expert labels, when available, are limited to recording-level summaries (i.e., one label per EEG recording). Hence, the applicability of supervised learning to discover sub-recording level (seconds to minutes) EEG patterns is currently limited.

On the other hand, traditional unsupervised learning approaches such as principal components analysis (PCA) and independent components analysis (ICA) require EEG data to be transformed into a two-dimensional matrix despite its natural three-dimensional structure. Furthermore, deep-learning-based approaches such as auto-encoders and self-supervision suffer from a lack of interpretability.

Unsupervised tensor decomposition (TD) is a natural choice for analyzing multi-dimensional EEG datasets\cite{cong2015tensor}. However, most prior research has focused on analyzing task-induced or stimulation-induced activity \cite{pouryazdian2017tensor}. TD has not been widely used for resting-state EEGs because of the lack of synchronization between segments. In this study, we build a population EEG tensor from large-scale resting-state clinical EEGs, transform the segments into the frequency domain to synchronize them, and decompose the tensor to discover meaningful patterns and investigate their interpretation and clinical utility.

Specifically, we utilize clinical EEGs from the TUH-EEG corpus to construct a single large EEG tensor with temporal, spatial, and spectral dimensions. This population-level EEG tensor is then decomposed to yield expert-interpretable patterns along those dimensions. We also conducted experiments to confirm the interpretation of the patterns and assess their ability to classify mild cognitive impairment (MCI) and Alzheimer's dementia (AD) against cognitively normal (CN).

Our study makes the following contributions:

\begin{itemize}
    \item We show that much of the variability in a population EEG dataset comprising multiple diseases can be sufficiently explained using only three latent factors. 
    \item Those parsimonious factors reflect clinically meaningful patterns and associate with known EEG correlates of cerebral function and signal artifacts. 
    \item Despite their parsimony, TD factors perform competitively in classifying stages of cognitive impairment compared to classical and deep-learning-based baselines that utilized large numbers of features (AUC = 0.92 for CN vs. AD and AUC = 0.65 for CN vs. MCI).
\end{itemize}

\section{Related Work}

The acquisition of brain-related bio-signals (functional or structural neuroimaging, magnetoencephalography, and EEG) produces inherently multi-dimensional datasets even in single-subject settings. Unsurprisingly, TD, which is a tool for higher-order data decomposition, has been extensively used to study such datasets. Broadly, TD has been used in brain-related research to 1) identify latent network or disease states \cite{cichocki2011tensor, leonardi2013identifying}, 2) extract features for discriminative analysis \cite{cichocki2011tensor,  zhang2020multi,leonardi2013identifying}, 3) analyze structural and functional brain connectomes \cite{mirzaei2019overlapping}, and 4) fuse multi-modal data such as simultaneous EEG-fMRI \cite{zhang2020multi}.

In the scalp EEG literature, the majority of TD-based studies have analyzed task-induced (also known as event-related potentials or ERPs) or stimulation-induced activity \cite{pouryazdian2017tensor}. A smaller number of studies have used resting-state EEGs for disease-specific analysis. These include frequency-dependent brain connectivity analysis of major depressive disorder subjects \cite{liu2022exploring}, decomposition of EEG oscillatory spectrum to characterize rehabilitation-related sensorimotor activity \cite{rovst2020comparison}, detection of epileptic spikes \cite{hunyadi2017tensor}, and sleep staging \cite{kouchaki2014tensor}.

In contrast to such prior work that focuses on small datasets and single disorders, we use a large population-level dataset comprising patients with multiple disorders and focus on evaluating the clinical value in an unseen dataset.

\section{Materials \& Methods}
\label{sec:data}
\begin{figure*}
\begin{center}
\includegraphics[width=\textwidth]{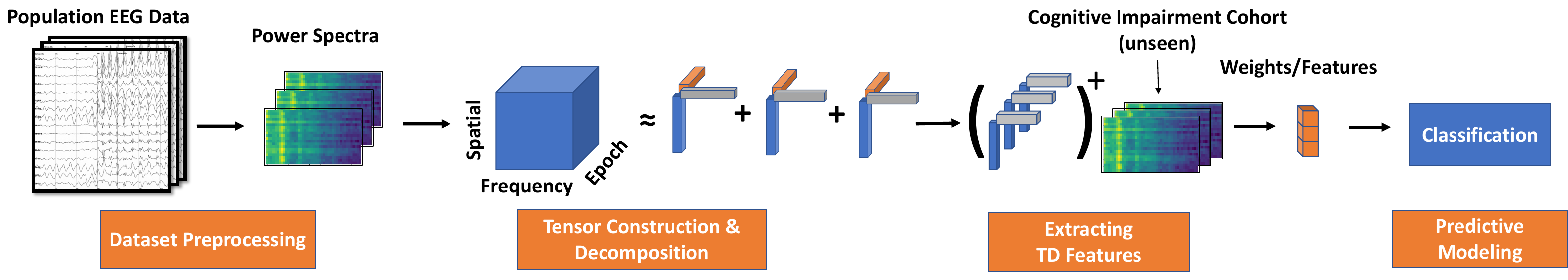}
\caption{Our overall analytic workflow. We construct a 3-D tensor using the power spectral densities of a population sample of eyes-closed awake EEG epochs and perform a tensor decomposition. Then, we obtain the weights for the EEG epochs of a validation cohort with respect to the discovered spatiospectral patterns using a projection algorithm and perform classification using those weights as features. 
}
\label{fig:workflow}
\end{center}
\end{figure*}

\noindent\textbf{Analytic workflow:} Figure \ref{fig:workflow} illustrates the overall workflow of our study. First, we preprocess raw EEG signals to extract spectral information and construct a population-level EEG tensor. Then, we decompose this tensor to discover multiple EEG patterns (Section \ref{sec:methods}). Each pattern is defined by a distinct signature in the spatial and spectral domains. Finally, we validate the clinical value of these population data-driven patterns using an expert-labeled unseen cohort of patients with varying stages of cognitive impairment (Section \ref{sec:exp_setup}).
\noindent\textbf{Dataset:} 
All data used in this study are from the TUH-EEG database \cite{obeid2016temple}. This dataset comprises $\approx$30,000 EEG recordings collected at TUH starting from 2002. All EEGs were recorded with the standard 10-20 scalp EEG spatial layout.
A subset of TUH-EEG, including 2,342 routine EEGs (also known as the TUH abnormal corpus) was used to construct the population-level tensor. Another non-overlapping subset of the larger TUH-EEG corpus, including routine EEGs of 24 CN, 31 MCI, and 50 AD patients, was selected for validation.
These diagnoses were extracted from the text reports accompanying the EEGs by a board-certified neurologist. The breakdown of the two datasets by demographic variables is shown in Table \ref{table:data}. A previously published tool was used to extract age and gender information from the text reports \cite{rawal2021score}.

\begin{table}[h!]
    \caption{Composition of the population dataset ($N=2,342$) and the expert-labeled validation cohort ($N=105$).}
    \label{tab:pop_data}
    \centering
    \resizebox{\linewidth}{!}{
\begin{tabular}{c|c|ccc|cccc}
\hline\hline
\multirow{2}{*}{\textbf{Dataset}} & \multirow{2}{*}{\textbf{Labels}} & \multicolumn{3}{c}{\textbf{Gender}} & \multicolumn{4}{c}{\textbf{Age}} \\
\cline{3-9}
& & M & F & N/A & 18-30 & 30-50 & 50-70 & $>$70 \\
\hline
Population & - & 1064 & 1243 & 35 & 425 & 839 & 795 & 283\\
\hline
\hline
\multirow{3}{*}{Validation} & CN & 10 & 13 & 1 & - & - & 13 & 11\\
                                & MCI & 12 & 18 & 1 & - & 2 & 21 & 8\\
                                & AD & 6 & 42 & 2 & - & 1 & 25 & 24\\                                
\hline\hline
\end{tabular}}
  \label{table:data}
\end{table}

\noindent\textbf{EEG processing:} We processed the EEGs as follows: a) we arranged channels of all EEGs in the same order, b) applied a bandpass filter between 0.5Hz and 45.0Hz, c) divided the recordings into contiguous non-overlapping epochs of 10-seconds,
d) identified and removed bad epochs when the total power in their `Cz' channel exceeded 2 standard deviations as calculated from statistics of each recording, e) identified $n$ eyes closed awake epochs from each recording ($n \in \{2,\dots, 6\}$) \cite{varatharajah2022quantitative}, and finally f) obtained the Fourier transform of each epoch between 1--45 Hz using the Welch method. 

\label{sec:methods}

\noindent\textbf{Tensor construction and decomposition:} We use $x \in \mathbb{R}^{S \times F}$ to denote the Fourier transform of a single EEG epoch with $S$ sensors at $F$ frequencies. Then, a multi-subject dataset including a total of $E$ epochs is represented by a third-order tensor $X \in \mathbb{R}^{E \times S \times F}$. Next, we apply the Canonical Polyadic Decomposition (CPD) \cite{hitchcock1927expression} with rank $r$ to decompose the population-level tensor $X$ as a sum of $r$ rank-1 tensors $X_i = e_i \otimes s_i \otimes f_i$ for $i\in \{1, \dots, r\}$.

\begin{equation}
\label{eq:td}
X \approx \sum_{i=1}^r X_{i} = \sum_{i=1}^r e_i \otimes s_i \otimes f_i
\end{equation}
 
The vectors $e_i$, $s_i$, and $f_i$ represent the population-level patterns in the $E$, $S$, and $F$ dimensions, respectively. Although the choice of rank $r$ is recognized to be an NP-hard problem, heuristic approaches based on empirical reconstruction errors exist. 
In this work, we use DIFFIT  \cite{timmerman2000three} to determine the decomposition rank. Given a tensor, DIFFIT sweeps over a range of different rank values and uses a ratio of the explained sum of squares to determine the ideal rank for decomposition. The optimization of the CPD for a given rank $r$ was performed using the Gauss-Newton approach \cite{paatero1997weighted}. 

\noindent\textbf{Extracting features for new data:}
 
Here we aim to represent new data in the space of these population-level spatiospectral patterns discovered using CPD, i.e., with respect to $s_i$ and $f_i$. This is done in two steps. 
First, we form a basis matrix $B \in \mathbb{R}^{S\cdot F \times r}$ for the space spanned by all $s$'s and $f$'s. Here, the columns of $B$ are vectorized versions of $(s_{i} \otimes f_{i})$, $i\in \{1, \dots, r\}$. Then, for a given epoch $x_{new} \in \mathbb{R}^{S \times F}$, we obtain the coordinates $w \in \mathbb{R}^{r}$ in the above space as $w = B^{+}\times\text{vec}(x_{new})$, where $\text{vec}(x_{new})$ represents vectorized $x_{new}$ and $B^{+}$ is the pseudo-inverse of $B$. Using this procedure, data from the validation cohort is projected onto the spatiospectral basis $B$ recovered from CPD.

\noindent\textbf{Predictive modeling:}
In the final step, we use the weight vector $w$ returned by projection as features to classify the validation cohort. We adopted a 15-fold cross-validation approach where the folds are made with disjoint sets of subjects. Although model training is done at the epoch level, the epoch-level predictions are averaged to obtain subject-level predictions. These subject-level predictions are then used to compute the area under the receiver operating characteristic curve (AUC) that summarizes model performance.

\section{Experiments \& Results}
\label{sec:exp_setup}

\begin{figure}[b!]
    \centering
    \subfigure[]{
    \includegraphics[width=0.33\textwidth]{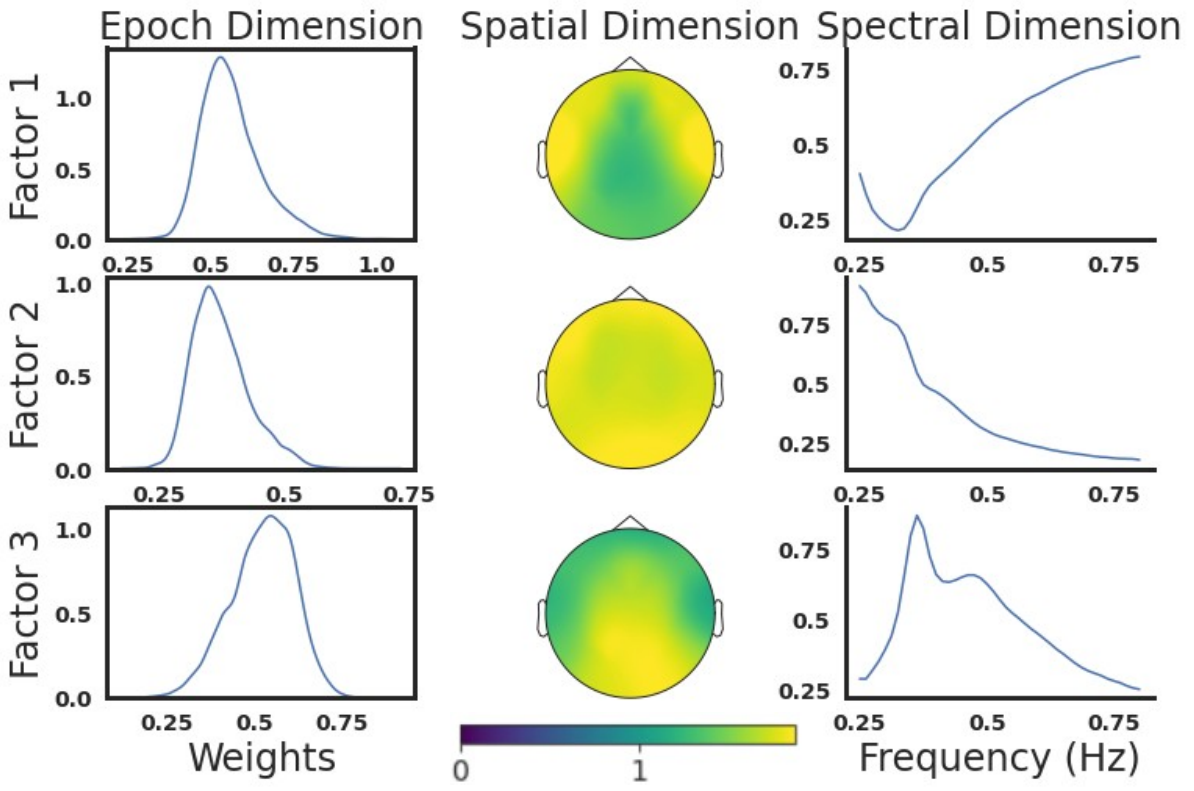}}
    \subfigure[]{
    \includegraphics[width=0.13\textwidth]{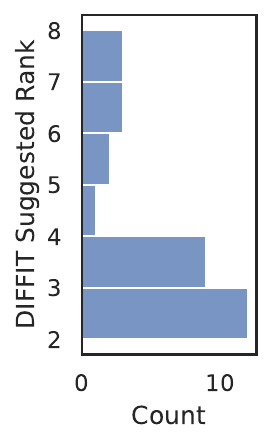}}
    \caption{(a) The features extracted from our population-level tensor. (b) Histogram of suggested rank over 30 runs of DIFFIT.
    }
    \centering
    \label{fig:factors}
\end{figure}

\begin{figure*}[t!]
    \centering
    \subfigure[]{\includegraphics[ width=0.17\textwidth]{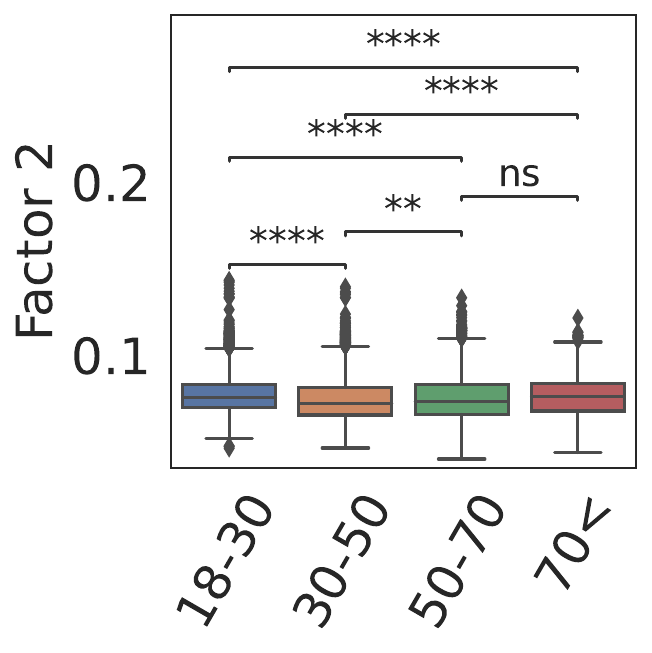}} 
    \subfigure[]{\includegraphics[ width=0.17\textwidth]{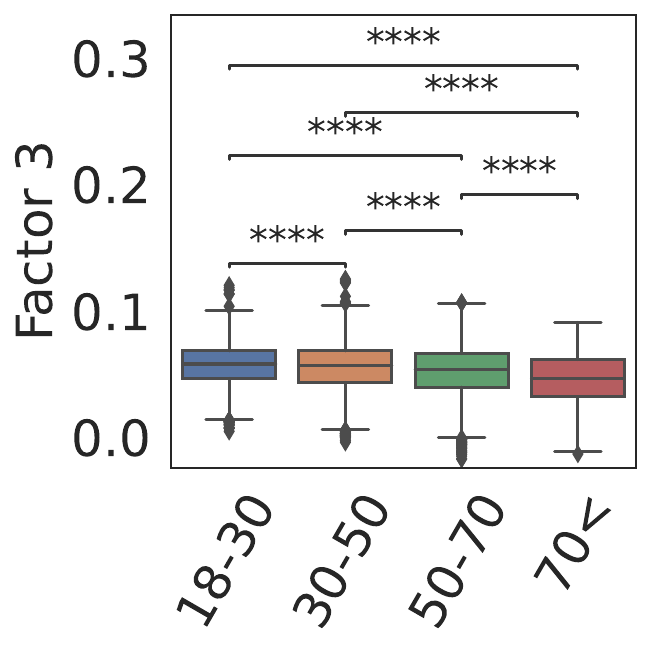}} 
    \subfigure[]{\includegraphics[ width=0.19\textwidth]{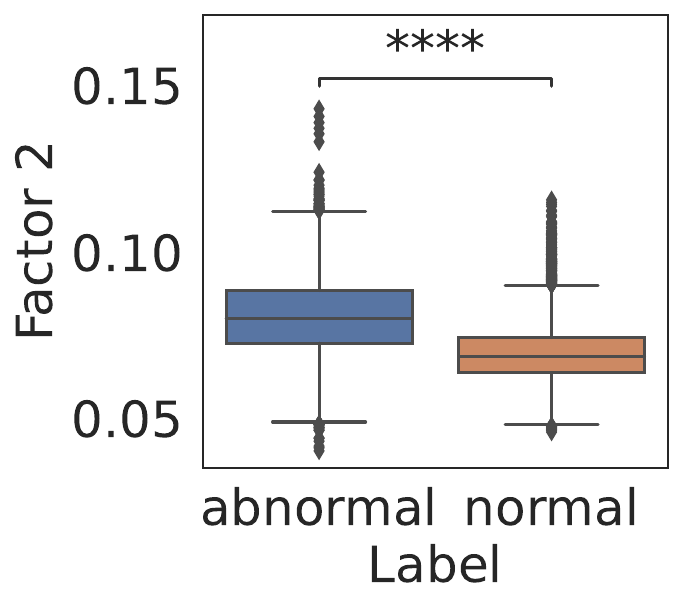}} 
    \subfigure[]{\includegraphics[ width=0.19\textwidth]{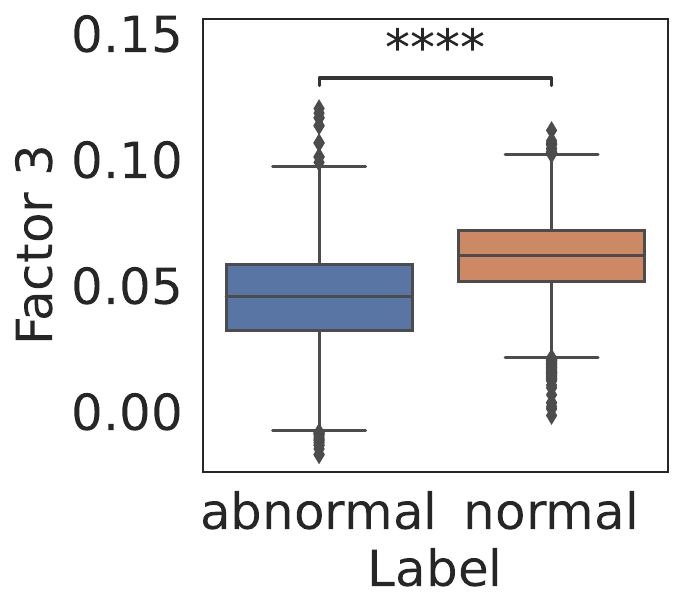}} 
    \subfigure[]{\includegraphics[ width=0.19\textwidth]{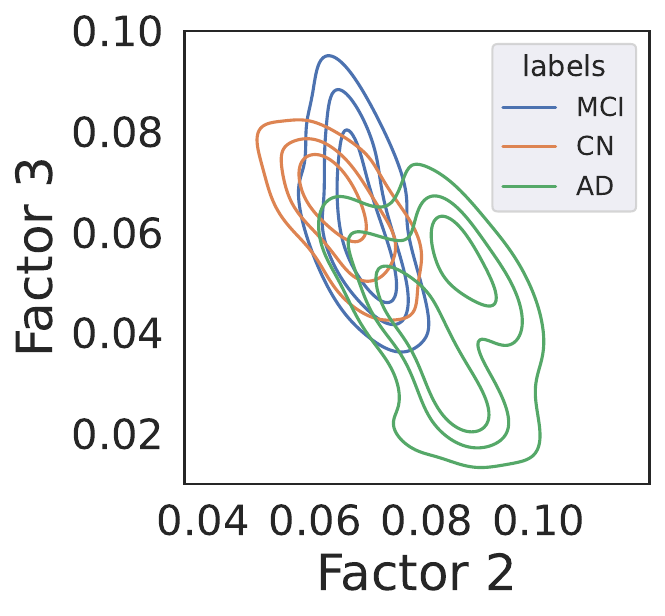}}

    \caption{(a-d) Boxplots of weights of factors 2 and 3 across multiple age groups and normal/abnormal EEG labels. (e) Kernel density plot of cognitive impairment groups over the weights of factors 2 and 3. Stars denote statistically significant differences with p-values 0.05 (*), 0.01 (**), 0.001 (***), and 0.0001 (****), and ns denotes not significant (all based on $t$-tests).}
    \label{fig:boxplots}
    \centering
\end{figure*}

We conduct experiments to interpret the factors and validate their utility in a clinical application. Our results are summarized in Figures \ref{fig:factors} and \ref{fig:boxplots}, and Table \ref{table:auc}.

\noindent\textbf{Parsimonious EEG patterns:}  
We obtained the optimal DIFFIT score for $r=3$, which results in a stable decomposition across multiple initializations, shown in Figure \ref{fig:factors}. In Figure \ref{fig:factors}b, which shows the number of times each rank was selected as optimal by DIFFIT, we see a sharp decrease in the counts after $r=3$. In the following, we report results using $r=3$. 
This result suggests that despite the complexity of scalp EEG, a large portion of the variability in population-level EEG can be explained by as few as three latent patterns.

\noindent\textbf{Relations with brain physiology:}
Figure \ref{fig:factors}a visualizes the rank-1 tensors representing the three factors along the epoch, space, and frequency dimensions. The rank-1 tensors in the epoch ($\in \mathbb{R}^E$), spatial ($\in \mathbb{R}^S$), and frequency ($\in \mathbb{R}^F$) dimensions are plotted as a probability density, a topographical map, and a power spectral density, respectively. The epoch dimension represents the weights of each EEG epoch for the corresponding spatiospectral pattern; i.e., it reflects how much of that pattern is present in a particular epoch.
Although the scales of the individual patterns are not meaningful, reconstruction (Eq. \ref{eq:td}) will yield an approximation of the original tensor.
Figure \ref{fig:boxplots} shows the distribution of these epoch-level `weights' across multiple age groups, normal/abnormal expert labels, and various stages of cognitive impairment. We find that factors 2 and 3 show marked differences based on aging, abnormal grade, and degree of cognitive impairment. Below, we interpret the factors based on the spatiospectral patterns.

\noindent\textbf{Factor 1:} We observe a bilateral frontotemporal activation with high power in the higher frequencies ($>$25 Hz). This factor most likely indicates muscle artifacts involving the frontalis (eye) and/or temporalis (facial/jaw) muscles, respectively.

\noindent\textbf{Factor 2:} We observe a component with higher activation in the lower frequency range ($<$10 Hz) and diffuse scalp distribution. Moreover, the presence of this factor is increased in abnormal EEGs (Figure \ref{fig:boxplots}c). Thus, this factor most likely indicates generalized slowing related to pathologies.

\noindent\textbf{Factor 3:} We observe central-to-posterior activation with peaks in the alpha (8-12Hz) and beta (13-25Hz) ranges. This factor shows a decreasing trend with age and is increased in normal EEGs (Figure \ref{fig:boxplots}b,d). We surmise that this factor represents combined posterior alpha and central beta oscillations.

\noindent\textbf{Classification results:} 
Next, we investigate the utility of these factors in classifying CN vs. MCI and CN vs. AD. We computed the weights for the validation data using the projection algorithm (Section \ref{sec:methods}). Those weights serve as features in classification. We compared the performance of those features against two other unsupervised features: 1) classical expert features of power in frequency bands (PIB), and 2) embeddings from a self-supervised feature encoder (SSL) \cite{wagh2021domain}. We also evaluated three classifiers, Gaussian Naive Bayes, a support vector machine, and a shallow neural network, for each feature. The results are shown in Table \ref{table:auc}.

\begin{table}[h!]
\caption{Cognitive impairment classification with different features (TD - tensor decomposition, PIB - power in bands, SSL - self-supervised) and classifiers (GNB - Gaussian Naive Bayes, SVM - support vector machine, NN - neural network). Mean AUC and standard deviation across subjects are reported. Bold values indicate the best performance per task.}
\begin{center}

\renewcommand{\arraystretch}{1.2}

\resizebox{\linewidth}{!}{\begin{tabular}{c|c|cc}

\cline{1-4}\hline\hline
\textbf{Feature-Classifier} &
\textbf{Feature Size} &
\multicolumn{2}{c}{\textbf{Classification Result (AUC)}} \\
\cline{3-4}
 & & MCI vs CN & AD vs CN \\
\hline
  TD-GNB & 3 &$0.65 \pm 0.33$ & $0.89 \pm 0.16$ \\
  TD-SVM  & 3 &$0.46 \pm 0.32$ & $\boldsymbol{0.92 \pm 0.15}$ \\
  TD-NN & 3 & $0.61 \pm 0.31$ & $0.62 \pm 0.35$ \\
\hline
  PIB-GNB  & 95 & $0.49 \pm 0.38$ & $\boldsymbol{0.92 \pm 0.15}$ \\
  PIB-SVM  & 95 & $0.65 \pm 0.35$ & $0.86 \pm 0.18$ \\
  PIB-NN  & 95 &$0.45 \pm 0.37$ & $0.76 \pm 0.30$ \\
  \hline
  SSL-GNB & 128 & $0.45 \pm 0.40$ & $0.69 \pm 0.33$ \\
  SSL-SVM  & 128 & $0.52 \pm 0.35$ & $0.66 \pm 0.28$ \\
  SSL-NN  & 128 & $\boldsymbol{0.69 \pm 0.30}$ & $0.68 \pm 0.29$ \\
 \cline{1-4}\hline\hline
\end{tabular}}
\label{table:auc}
\end{center}

\end{table}

We observe that most combinations of features and models perform poorly in classifying MCI vs. CN. Self-supervised features perform the best (AUC: 0.69). Interestingly, TD features, although much fewer in number, turn out to be a close second (AUC: 0.65). 
In AD vs. CN, we find that TD-SVM and PIB-GNB both perform the best (AUC: 0.92), followed closely by TD-GNB (AUC: 0.89) and PIB-SVM (AUC: 0.86).

\section{Discussion \& Future Work}

This study demonstrates the use of unsupervised tensor decomposition to recover interpretable EEG patterns from population-level clinical EEGs in a purely data-driven manner. Furthermore, we show their clinical value by classifying a cohort of patients with various stages of cognitive impairment.
\noindent\textbf{Significance:}
The increasing availability of large EEG datasets and the laborious and error-prone nature of expert EEG annotations have created a need for unsupervised or weakly supervised approaches that can extract clinically interpretable features from large amounts of unlabeled data. This paper addresses this need by highlighting the utility of tensor decomposition to recover population-level EEG patterns in a data-driven manner. Our results suggest that these EEG patterns correspond to known brain pathophysiology and are useful in disease diagnosis tasks involving cognitive impairment.
\noindent\textbf{Limitations and future directions:} 
Here we used an algorithmic approach to determine the rank (i.e., \# features) in TD. A more principled approach guided by domain knowledge can recover more subtle EEG patterns specific to diseases. Next, additional out-of-sample evaluations using data from various disease groups, such as epilepsy, stroke, and psychiatric disorders, can further validate the usefulness of TD and the identified factors. Finally, the same analytic workflow can be applied to functional/spectral connectivity data to recover whole-brain network motifs and aid in discovering network-based patterns at the population level.

\section{conclusion}
We proposed a tensor decomposition-based approach to extract meaningful features from population-level EEG data. We performed experiments using the large-scale TUH-EEG dataset and an expert-labeled cohort of patients including varying degrees of cognitive impairment. Our approach discovered a parsimonious set of EEG patterns ($r=3$) from large population-level data ($N = 2,342$ EEGs). Our results suggest that these EEG patterns correspond to known brain physiology and are useful in classifying different stages of cognitive impairment. Future studies are needed to evaluate the proposed approach in diagnosing other neurological diseases.
\vspace{2mm}
\noindent\textbf{Acknowledgements:} This research was partially supported by the National Science Foundation (Award No. IIS-2105233) and a Mayo Clinic \& Illinois Alliance Fellowship.

\bibliographystyle{IEEEtran}
\bibliography{IEEEabrv,references}

\end{document}